\documentclass[preprint2]{aastex61}
\usepackage{amsmath}
\usepackage{amssymb}

\begin{document}

\title{Overshooting in the core helium burning stage of a $30M_{\odot}$ star \\
using the $\MakeLowercase k$-$\omega$ model}

\correspondingauthor{Yan Li}
\email{ly@ynao.ac.cn}

\author{Yan Li}
\affiliation{Yunnan Observatories, Chinese Academy of Sciences, Kunming 650216, China}
\affiliation{Key Laboratory for the Structure and Evolution of Celestial Objects, Chinese Academy of Sciences}
\affiliation{Center for Astronomical Mega-Science, Chinese Academy of Sciences, Beijing 100012, China}
\affiliation{University of Chinese Academy of Sciences, Beijing 100049, China}

\author{Xing-hao Chen}
\affiliation{Yunnan Observatories, Chinese Academy of Sciences, Kunming 650216, China}
\affiliation{Key Laboratory for the Structure and Evolution of Celestial Objects, Chinese Academy of Sciences}

\author{Hai-liang Chen}
\affiliation{Yunnan Observatories, Chinese Academy of Sciences, Kunming 650216, China}
\affiliation{Key Laboratory for the Structure and Evolution of Celestial Objects, Chinese Academy of Sciences}


\begin{abstract}
Overshooting and semiconvection are among the most uncertainties in the evolution of massive stars. Complete mixing over a certain distance beyond the convective boundary (Stothers \& Chin 1985) and an exponentially decaying diffusion outside the convection zone (Herwig 2000) are commonly adopted for the overshoot mixing. Recently, Li (2012, 2017) developed the $k$-$\omega$ model, which can be used in both convection zones and overshooting regions. We incorporated it in calculations of $30M_{\odot}$ stellar models. We find that in the main sequence stage, models with the $k$-$\omega$ model are almost identical to models with complete mixing in the overshooting region beyond the convective core, and the overshooting in the $k$-$\omega$ model is equivalent to an overshooting distance of about $0.15H_P$. In the post main sequence stage, we find that the overshooting below the bottom of the intermediate convection zone beyond the hydrogen-burning shell can significantly restrict the size of the hydrogen-depleted core, and can penetrate effectively into the hydrogen-burning shell. These two effects are crucial for the evolution of the core helium burning stage. During the core helium burning stage, we find that the overshooting model based on the $k$-$\omega$ model results in a similar complete mixing region but a much wider partial mixing region than the overshooting model based on Herwig (2000). In particular, the overshooting distance in the core helium burning stage may be significantly smaller than that in the main sequence phase for massive stars.
\end{abstract}

\keywords{stars: massive --- 
          stars: evolution --- 
          stars: interiors  --- 
          convection}

\section{Introduction}

Overshooting and semiconvection are among the most uncertainties in the evolution of massive stars (Woosley et al. 2002; Langer 2012). During the main sequence stage, massive stars will form convective cores. Many investigations indicate that overshooting should be present beyond the convective core (Stothers \& Chin 1985; Briquet et al. 2007; Hunter et al. 2008; Brott et al. 2011). The overshoot mixing may help to explain the width of the main sequence on the Hertzsprung-Russell diagram (Stothers \& Chin 1985). Near and after the end of the main sequence stage, semiconvection can be developed in the chemical gradient region (Schwarzschild \& H\"arm 1958; Chiosi \& Summa 1970; Langer, El Eid \& Fricke 1985). As pointed out by Kato (1966), semiconvection is a kind of overstable oscillation with its amplitude increasing exponentially to finally result in mixing. Recently, Spruit (1992, 2013) modelled semiconvection as a double-diffusive layer, where many thin layers are formed to allow mixing between those of convection and diffusion. Mixing in the semiconvection zone is crucial for the evolution of massive stars: a strong mixing will result in helium to be ignited at the stellar center in the blue supergiant phase, while a weak mixing will cause helium to ignite in the red supergiant phase (Langer \& Maeder 1995; Maeder \& Meynet 2001). During the core helium burning stage, convective cores are formed again, and the overshooting beyond the convective core plays a critical role in determining the size of the final C/O core (Woosley \& Weaver 1988). 

There are two kinds of approaches to be commonly adopted to treat the overshoot mixing in the literature. Firstly, the overshoot mixing is applied in stellar models by directly increasing the convective region by a certain amount (e.g. Stothers \& Chin 1985). The overshooting distance is usually assumed to be proportional to the local pressure scale height, and the coefficient is considered as a free parameter to be determined by comparisons with observations (Mermilliod \& Maeder 1986; Maeder \& Meynet 1987; Ekstr\"om et al. 2008; Briquet et al. 2007; Hunter et al. 2008; Brott et al. 2011). The other is based on numerical simulations of stellar convection (Freytag et al. 1996), and describes the overshooting region by a diffusion approach with an exponentially decaying diffusion coefficient (e.g. Herwig 2000). Similarly, a model parameter $f_{\rm OV}$ should be determined by comparisons with observations. 

Calibrations of overshooting parameters are essential for applications of any overshooting formalisms. Based on the main sequence widths of star clusters, Mermilliod \& Maeder (1986) and Maeder \& Meynet (1987) obtained an overshooting of $0.23\sim 0.3H_P$ from comparisons with nonrotating stellar models, and Ekstr\"om et al. (2008) found an overshooting of $0.1H_P$ for rotating stellar models. Based on asteroseismological investigations, Briquet et al. (2007) determined the overshooting to $0.44H_P$ for the $\beta$-Cephei variable $\theta$ Ophiuchi, whereas Moravveji et al. (2015) found the overshooting to $f{\rm ov}\approx 0.017$ for a B8.3V star KIC 10526294. However, overshooting in the core helium burning stage has not yet been reasonably estimated so far for massive stars.

Various approaches have been proposed to treat the mixing in the semiconvection region. Langer et al. (1983) proposed an approximate numerical model based on the linear analysis of Kato (1966). Based on the double diffusive picture, Spruit (1992, 2013) approximated the mixing in the semiconvection layer by a semiconvective diffusion coefficient. On the other hand, the timescale of semiconvection is usually considered to be the thermal timescale, which is much shorter than the nuclear timescale in the main sequence stage. As a result, the detailed treatment of semiconvection makes little difference for the main-sequence evolution, and a complete mixing with the Schwarzschild criterion for a convection zone is often adopted in models of massive stars.

As originally pointed out by Xiong (1978, 1985), stellar convection is going to develop into turbulence. Therefore, the turbulent convection models (Xiong 1985; Xiong et al. 1997; Canotu 1992, 1993; Li \& Yang 2001), which are fully based on moment equations of fluid hydrodynamics, are appropriate to be used to describe the convective overshooting. Using the turbulent convection models, the effects of overshooting on the main sequence evolution of massive stars have been studied (Xiong 1986; Ding \& Li 2014a, 2014b). 

Recently, Li (2012, 2017) proposed a $k$-$\omega$ model for stellar convection. The $k$-$\omega$ model is based on moment equations of fluid hydrodynamics. Under appropriate assumptions, generations due to buoyancy and the shear from convective rolling cells and dissipation of turbulent kinetic energy are reasonably included, and the turbulent transport effect is approximated by a gradient-type diffusion model (see Section 2 for details). It has been applied already in the solar models, sdB models, and evolution of low and intermediate mass stars (Li 2017; Li et al. 2018). The $k$-$\omega$ model includes eight free parameters (Li 2017), which should be determined in advance for applications in stars. Unlike those traditional overshooting formalisms that are phenomenological prescription (e.g., Stothers \& Chin 1985) or fitting formula to direct numerical simulations (e.g., Herwig 2000), the $k$-$\omega$ model is a fully dynamic theory for stellar turbulent convection. It can approximately describe the turbulent motions by two differential equations, directly resulting in the properties of turbulent convection in stars. One advantage of this property is that it can be applied to similar situations when its parameters has been properly determined in advance. For example, it can be used to the convective core during the core helium burning stage, if its parameters has been determined for the convective core of a star during the main sequence. This provides us an opportunity to explore the overshooting in the core helium burning stage of massive stars, with the aid of observational constraints in the main sequence. 

In the present paper, we show the applications of the $k$-$\omega$ model in the evolution of a $30M_{\odot}$ star, not only in the main sequence but also in the central helium burning stages. In the main sequence stage, we compare stellar models of the $k$-$\omega$ model with stellar models of other overshooting formalisms, in order to calibrate the free parameters in the $k$-$\omega$ model under constraints of observations. Then we continue to evolve the stellar models to core helium burning stage, to calibrate the overshooting that cannot be properly constrained up to now both theoretically and observationally. In Section 2, we introduce equations of the $k$-$\omega$ model. In Section 3, we briefly summarize the input physics of our calculations of stellar models, and introduce how to incorporate the $k$-$\omega$ model into our stellar models. Our main results are discussed in Section 4 for stellar models in the main sequence stage and in Section 5 for stellar models in the central helium burning stage. In Section 6, we summarize our main conclusions.

\section{ Equations of the $\MakeLowercase{k}$-$\omega$ model for stellar convection }

Because of great dimensions of stellar convection zones, the stellar convective motions must evolve into fully developed turbulence and should be described by turbulence models. We describe the stellar turbulent convection by a $k$-$\omega$ model developed by Li (2012, 2017), which is fully based on moment equations of fluid hydrodynamics. The turbulent convection is described by an equation for turbulent kinetic energy $k$:
\begin{equation}\label{2c1}
\frac{\partial k}{\partial t}-\frac{1}{{{r}^{2}}}\frac{\partial }{\partial r}\left( {{r}^{2}}{{\nu }_{t}}\frac{\partial k}{\partial r} \right)=S+G-k\omega ,
\end{equation}
and an equation for turbulence frequency $\omega$:
\begin{equation}\label{2c2}
\frac{\partial \omega }{\partial t}-\frac{1}{{{r}^{2}}}\frac{\partial }{\partial r}\left( {{r}^{2}}\frac{{{\nu }_{t}}}{{{\sigma }_{\omega }}}\frac{\partial \omega }{\partial r} \right)=\frac{c_{L}^{2}k}{{{L}^{2}}}-{{\omega }^{2}},
\end{equation}
where $S$ and $G$ respectively represent the shear and buoyancy production rate of the turbulent kinetic energy, $L$ represents the macro-length of turbulence that is equivalent to the mixing-length in the standard mixing-length theory. In Equations (\ref{2c1}) and (\ref{2c2}), the turbulent diffusivity $\nu_{t}$ is approximated by:
\begin{equation}\label{2c3}
{{\nu }_{t}}={{c}_{\mu }}\frac{k}{\omega }.
\end{equation}
According to common choices (e.g., Pope 2000), the model parameters $c_{\mu}=0.09$ and ${c}_{L}=c_{\mu }^{3/4}$, and $\sigma_{\omega}=1.5$.

According to Li (2012, 2017), the buoyancy production rate is approximated by
\begin{equation}\label{2c4}
G=-\frac{{{c}_{t}}}{1+\frac{\lambda \omega }{\rho {{c}_{P}}k}+{{c}_{t}}{{c}_{\theta }}{{\omega }^{-2}}{{N}^{2}}}\frac{k}{\omega }{{N}^{2}},
\end{equation}
where the buoyancy frequency $N$ is defined in a chemically homogeneous region as:
\begin{equation}\label{2c5}
{{N}^{2}}=-\frac{\beta gT}{{{H}_{P}}}\left( \nabla -{{\nabla }_{ad}} \right),
\end{equation}
and the radiation diffusivity $\lambda$ is defined as:
\begin{equation}\label{2c6}
\lambda =\frac{16\sigma {{T}^{3}}}{3\rho \kappa }.
\end{equation}
In above equations, $\rho$ is the density, $P$ the pressure, $T$ the temperature, ${{c}_{P}}$ the specific heat at constant pressure, $g$ the gravitational acceleration, $H_P$ the local pressure scale height, $\kappa$ the Rosseland mean opacity, $\sigma$ the Stefan-Boltzmann constant, $\nabla$ and $\nabla_{ad}$ are respectively the actual and adiabatic temperature gradient, and the thermodynamic coefficient $\beta$ is defined by
\begin{equation}\label{2c7}
\beta =-\frac{1}{\rho }{{\left( \frac{\partial \rho }{\partial T} \right)}_{P}}.
\end{equation}
$c_t$ and $c_{\theta}$ in Equation (\ref{2c4}) are two parameters, and their values are given by turbulence models (e.g., Hossain \& Rodi 1982): $c_t=0.1$ and $c_{\theta}=0.5$. 

The shear production rate is approximated by
\begin{equation}\label{2c8}
S=\frac{{{c}_{\mu }}}{{{c}_{L}}}\frac{q}{1+{{q}^{2}}}k\omega ,
\end{equation}
where $q$ is defined by
\begin{equation}\label{2c9}
q={{\left( \frac{\rho {{c}_{P}}{{L}^{2}}}{\lambda }\omega  \right)}^{-2/3}}.
\end{equation}

The macro-length of turbulence $L$ is usually assumed to be proportional to the local pressure scale height $H_P$, if the thickness of the stellar convection zone is much greater than the local pressure scale height: 
\begin{equation}\label{2c10}
L={{c}_{L}}\alpha {{H}_{P}},
\end{equation}
where $\alpha$ is a model parameter. For convective cores, however, their sizes are usually smaller than the local pressure scale height. Accordingly, the macro-length of turbulence $L$ will be restricted by the actual size of the convection core. We therefore adopt the model of \citet{ly17} for the macro-length of turbulence in the convective core:
\begin{equation}\label{2c11}
L={{c}_{L}}{\alpha }'{{R}_{cc}},
\end{equation}
where $R_{cc}$ is the radius of the convective core, and ${\alpha }'$ is a model parameter.

The evolution of element abundance in the stellar interior can be approximately treated by a diffusion equation:
\begin{equation}\label{2c12}
\frac{\partial {{X}_{i}}}{\partial t}=\frac{\partial }{\partial m}\left[ {{\left( 4\pi \rho {{r}^{2}} \right)}^{2}}\left( {{D}_{mix}}+{{D}_{t}} \right)\frac{\partial {{X}_{i}}}{\partial m} \right]+{{d}_{i}},
\end{equation}
where $X_i$ is the mass fraction and $d_i$ the generation rate of element “$i$”. $D_{mix}$ represents the diffusion coefficient due to any mixing process, such as element diffusion, meridional circulation, etc., while $D_t$ is due to the convective and/or overshoot mixing and is approximated by:
\begin{equation}\label{2c13}
{{D}_{t}}=\frac{{{c}_{X}}}{1+\frac{\lambda \omega }{\rho {{c}_{P}}k}+{{c}_{t}}{{c}_{\theta }}{{\omega }^{-2}}{{N}^{2}}}\frac{k}{\omega }.
\end{equation}
In Equation (\ref{2c13}), $c_X$ is a model parameter.

\section{ Input physics and numerical schemes }

Our stellar models are computed with the Modules of Experiments in Stellar Astrophysics (MESA), which was developed by Paxton et al. (2011, 2013, 2015, 2018). We use version 6596 of MESA and choose the package of "7M\_prems\_to\_AGB". The detailed input physics is described as follows. We consider the evolution of star with an initial mass of $30M_{\odot}$. We choose the initial composition of our stellar models to be $(Y,Z)=(0.28,0.02)$. As an example to explore the overshooting in the core helium burning stage, it is suitable to discuss the stellar evolution with only one value of the initial mass and chemical composition. We have calculated the evolution of the star from the pre-main-sequence to the depletion of helium at the stellar center. We use the "basic" nuclear network of MESA, which include pp chains and CNO cycles for hydrogen burning and helium burning reactions. In order to evolve the stellar models to the helium burning stage, we use the OPAL opacity data with GN93 composition (Grevesse \& Noels 1993), which include extra C and O enhancement. The Eddington grey atmosphere is used as the outer boundary condition. We adopt "Dutch" scheme of MESA for the mass loss due to stellar wind with the wind coefficient $\eta=1.0$, which is exactly the same as that used in Choi et al. (2016). 

We use the $k$-$\omega$ model to treat convection, whose parameters are summarized in Table 1. It should be noted that the values of the adjustable parameters ($c_X$, $\alpha$, and $\alpha'$) are adopted according to solar and stellar calibrations (Li 2012, 2017). The Schwarzschild criterion is adopted to determine the convection zone. We also use the standard mixing-length theory to calculate stellar models for comparisons, using the mixing-length parameter $\alpha=1.73$. Mixing due to rotation is not included in our stellar models, though it is believed to be significant in massive stars (Maeder 1987; Langer 1991; Maeder \& Meynet 2000; Maeder 2009). 

\begin{deluxetable}{cccccccc}
\tablewidth{0pt}
\tablecaption{Parameters of the $k$-$\omega$ model}
\tablehead{
\colhead{$c_{\mu}$}    & \colhead{$\sigma_{\omega} $} & 
\colhead{$c_{\theta}$} & \colhead{$c_t$} & 
\colhead{$c_h$}        & \colhead{$c_X$} &        
\colhead{$\alpha$}     & \colhead{${\alpha}'$}  
}
\startdata
0.09 & 1.5 & 0.5 & 0.1 & 2.344 & 0.01 & 0.7 & 0.06 \\
\enddata
\end{deluxetable}

We have implemented the $k$-$\omega$ model into MESA v6596 through “run\_star\_extras.f” package as follows. We use the $k$-$\omega$ model for every convection zone. It should be noted that the way we have used to treat a convective core is slightly different from the way to treat an intermediate convection zone. For a convective core, we use Equation (\ref{2c11}) for the macro-length of turbulence $L$, while for an intermediate convection zone we use Equation (\ref{2c10}). When a stellar model is obtained by solving the equations of the stellar structure, we solve Equations (\ref{2c1}) and (\ref{2c2}) in the “run\_star\_extras.f” package to obtain $D_t$. Then we apply $D_t$ to continue calculations of the chemical evolution with MESA.

\section{Properties of convection during the main sequence stage}

During the central hydrogen burning phase, the star of $30M_{\odot}$ develops a convective core. The overshoot mixing beyond the convective core is considered in two types: according to the $k$-$\omega$ model and according to Herwig (2000). We adopt the overshooting parameter $f_{\rm OV}=0.012$ in Herwig (2000) approach, in order to match its result to that of the $k$-$\omega$ model. The mass enclosed inside the convective core reduces with the hydrogen consumption in the convective core, leaving a chemical gradient region behind. When the central hydrogen abundance is below about 0.5, a convective shell is formed according to the Schwarzschild criterion around the upper edge of the chemical gradient region and becomes wider and wider. As this convective shell can be developed into the chemical gradient region, mixing in it is still uncertain. Either complete mixing (Stothers \& Chin 1985) or partial mixing due to semiconvection (Langer et al. 1983) is commonly adopted in the literature. We also include these two types of mixing in our stellar models, and consider three types of their combinations: overshooting of Herwig (2000) beyond the convective core plus complete mixing in the semiconvection zone (referred to as “OV+Sch”), overshooting of the $k$-$\omega$ model beyond the convective core plus complete mixing in the semiconvection zone (referred to as “KO+Sch”), and overshooting of the $k$-$\omega$ model beyond the convective core plus partial mixing in the semiconvection zone (referred to as “KO+Semi”). We adopt a diffusion coefficient of $2\times 10^6 {\rm cm}^2{\rm s}^{-1}$ for the partial mixing in the semiconvection zone, the reason will be discussed in Section 4.2.

\begin{figure}
\plotone{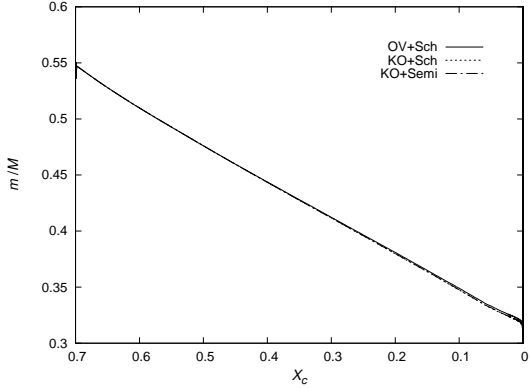}
\figurenum{1}
\caption{Development of the convective core during the main sequence stage. The abscissa is the central hydrogen abundance, while the ordinate is the mass fraction of the convective core. }
\end{figure}

\begin{figure}
\plotone{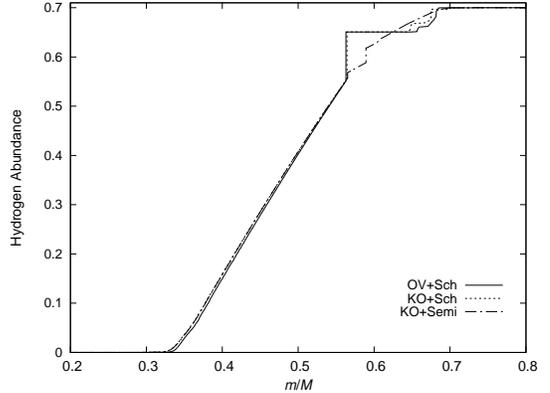}
\figurenum{2}
\caption{Hydrogen profiles at the end of the main sequence stage for the three considered stellar models.}
\end{figure}

The development of the convective core is shown in Figure 1, which displays the boundary of the convective core as a function of the central hydrogen abundance. It can be noticed that all of the three considered models result in almost identical convective cores. This result implies that the overshoot mixing of the $k$-$\omega$ model is roughly equivalent to that of Herwig (2000) with $f_{\rm OV}=0.012$. It should be noted that different treatments of mixing in the semiconvection region have almost no effect on the development of the convective core. The hydrogen profiles are shown in Figure 2 for those stellar models at the end of the central hydrogen burning phase. It can be seen that the hydrogen profiles are quite similar for KO+Sch and OV+Sch models, but fairly different in the region of $m/M>0.5$ for KO+Semi model due to different mixing schemes in the semiconvection zone.

\begin{figure}
\plotone{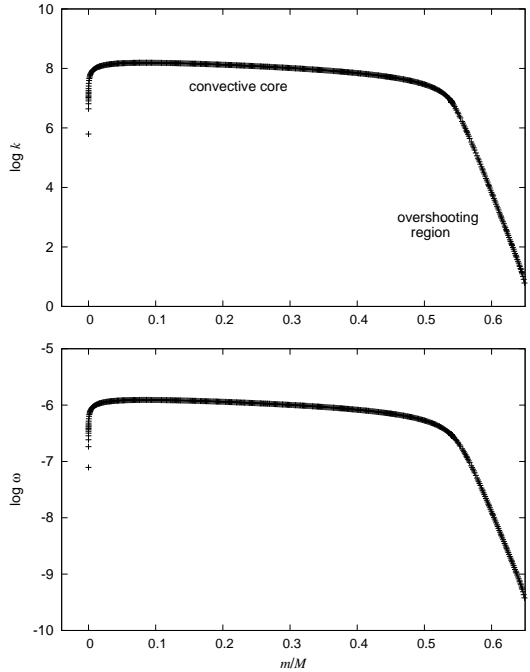}
\figurenum{3}
\caption{Distributions of the turbulent kinetic energy (upper panel) and the turbulence frequency (lower panel) in the convective core and the overshooting region for a KO+Semi model in the early main sequence stage. }
\end{figure}

\begin{figure}
\plotone{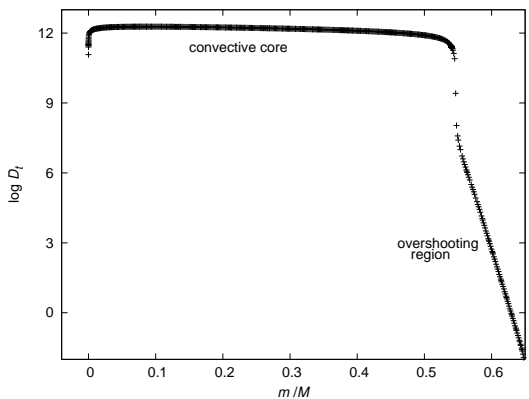}
\figurenum{4}
\caption{Distribution of the turbulent diffusivity in the convective core and the overshooting region for the same KO+Semi model in Fig. 3.}
\end{figure}

\subsection{ The convective core }

The distributions of the turbulent kinetic energy and turbulence frequency are shown in Figure 3 for a KO+Semi model in the early main sequence stage. It can be seen that the turbulent kinetic energy keeps at high values in the convective core, and decays exponentially in the overshooting region. It should be noted that the turbulent kinetic energy drops considerably near the stellar center, because it is inversely proportional to the local pressure scale height that is going to be infinity at the stellar center. The turbulent diffusivity is also shown for the same KO+Semi model in Figure 4. Again, it takes a high value to ensure complete mixing in the convective core, and decays exponentially to give a partial mixing in the overshooting region.

\begin{figure}
\plotone{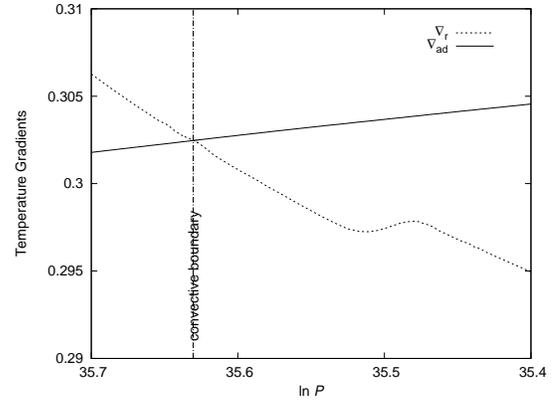}
\figurenum{5}
\caption{Distributions of both radiative and adiabatic temperature gradients for a stellar model in the early main sequence stage. The vertical line indicates the Schwarzschild boundary of the convective core.}
\end{figure}

\begin{figure}
\plotone{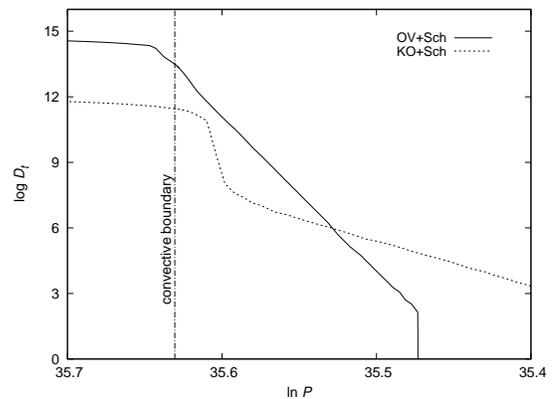}
\figurenum{6}
\caption{Distributions of the turbulent diffusivity near the boundary of the convective core for a OV+Sch model and a KO+Sch model in the early main sequence. The vertical line indicates the Schwarzschild boundary of the convective core. }
\end{figure}

In order to compare the properties of the overshooting region, we present the distributions of temperature gradients in Figure 5, and show the turbulent diffusivity for the OV+Sch model and the KO+Sch model in Figure 6. There are two features that should be noticed in Figure 6. The KO+Sch model results in a sharp decrease in the turbulent diffusivity near the boundary of the convective core, which will give a shorter complete mixing overshooting region beyond the convective core than the OV+Sch model does. On the other hand, the KO+Sch model results in a slower decay of the turbulent diffusivity, which will produce a longer partial mixing overshooting region than the OV+Sch model does. Figure 7 shows the hydrogen profiles of the two considered models. It can be seen that the whole overshooting region of the KO+Sch model is about $0.2H_P$, while that of the OV+Sch model is about $0.15H_P$. However, the complete mixing overshooting region of the KO+Sch model is only about half of that of the OV+Sch model. Our results of the overshooting distance is in agreement with observational constraints (Mermilliod \& Maeder 1986; Maeder \& Maynet 1987; Ekstr\"om et al. 2008; Moravveji et al. 2015).

\begin{figure}
\plotone{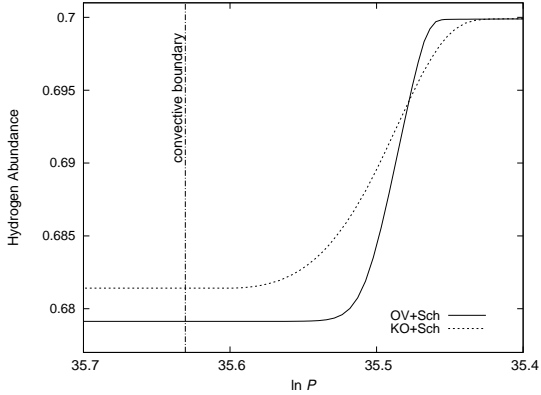}
\figurenum{7}
\caption{Distributions of the hydrogen abundance near the boundary of the convective core for both stellar models in Fig. 6. The vertical line indicates the Schwarzschild boundary of the convective core.}
\end{figure}

\begin{figure*}
\plotone{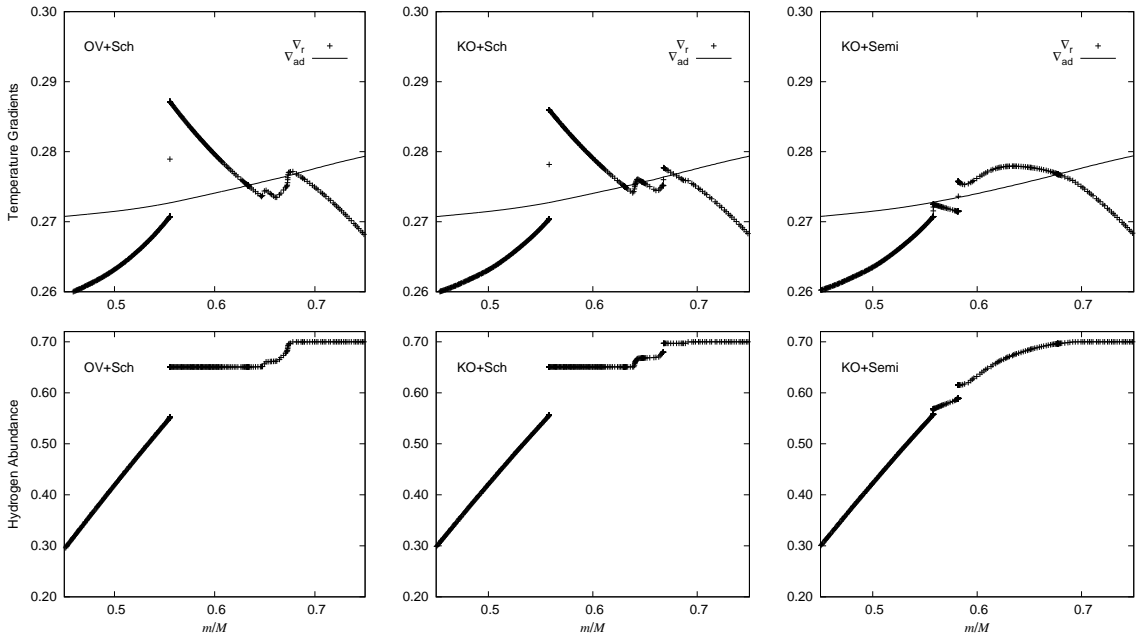}
\figurenum{8}
\caption{Distributions of temperature gradients (upper panel) and profiles of hydrogen abundance (lower panel) for the semiconvection zone of the three considered stellar models in the late main sequence stage. }
\end{figure*}

\subsection{ The semiconvection region }

When a convective shell appears near the upper edge of the chemical gradient region, mixing in it may become quite uncertain, depending on which stability criterion to be applied (Schwarzschild criterion or Ledoux criterion) and which mixing scheme to be used (complete mixing or partial mixing). The complete mixing is the most straightforward way, but it will bring about problematic results as shown in the left panels of Figure 8. It can be noticed that after application of the complete mixing in a convective region unstable against the Schwarzschild criterion, some part of it may become stable against the Schwarzschild criterion. Furthermore, continuous applications of complete mixing according to the Schwarzschild criterion would make many small step shapes on the hydrogen profile and many small burrs on the radiative temperature gradient in the same region. By comparing left and middle panels of Figure 8, the results of mixing sensitively depend on detailed practice of the mixing process itself, and a tiny difference at the beginning may result in a significant difference at the end.

Therefore, we adopt the partial mixing in the semiconvection region. We use a diffusion equation to describe the partial mixing, and choose a proper diffusion coefficient to keep the radiative temperature gradient always larger than the adiabatic temperature gradient in the whole semiconvection zone. In our calculations, we adopt the diffusion coefficient $D_{\rm mix}$ to be $2\times 10^6{\rm cm}^2{\rm s}^{-1}$. It can be seen in the right panels of Figure 8 that the semiconvection region always satisfies the Schwarzschild criterion, and the hydrogen profile keeps smooth as much as possible.

\begin{figure}
\plotone{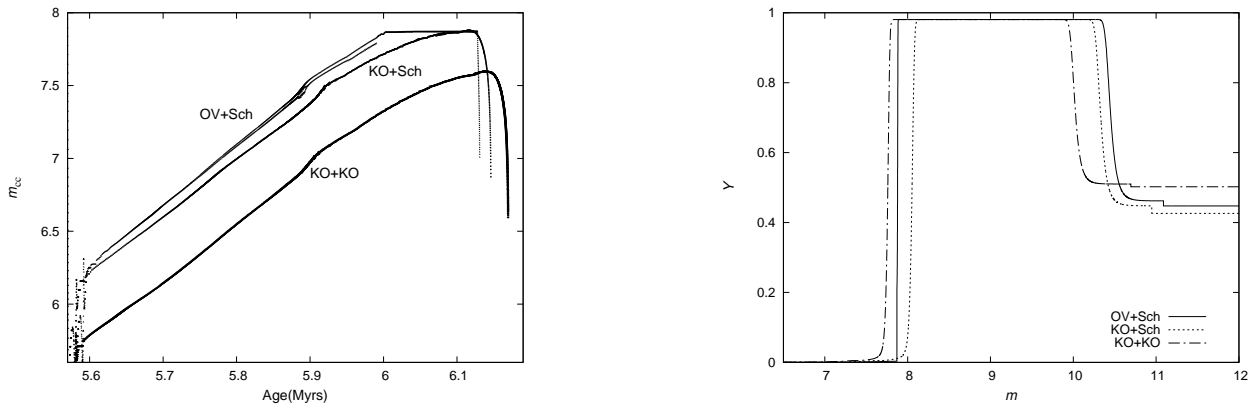}
\figurenum{9}
\caption{Development of the convective core during the central helium burning stage. The abscissa is the age in unit of Myrs, while the ordinate is the mass of the convective core in solar mass unit. }
\end{figure}

\begin{figure}
\plotone{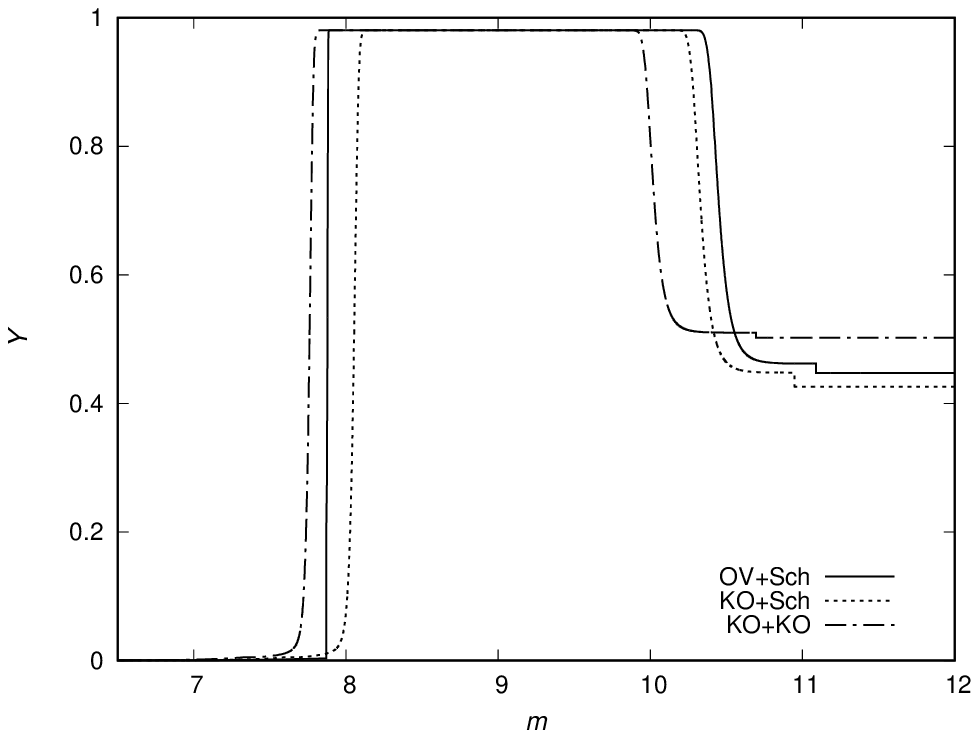}
\figurenum{10}
\caption{Helium profiles when helium is exhausted in the convective core for the three considered stellar models. The abscissa is the mass in unit of solar mass.}
\end{figure}

\section{Properties of convection during the helium burning stage}

As soon as hydrogen is depleted in the convective core, a helium core is formed. The helium core begins to contract and then to heat up, resulting in the formation of a hydrogen-burning shell on its surface. At about the same moment, an intermediate convection zone is formed not far above the hydrogen-burning shell, and soon expands to almost the entire chemical gradient region. We also consider two types of mixing in this intermediate convection zone: complete mixing inside the Schwarzschild boundaries and overshoot mixing according to the $k$-$\omega$ model. Soon later, helium ignites at the stellar center, leading to the formation of a new convective core. Afterwards, the convective core expands while the intermediate convection zone above the hydrogen-burning shell shrinks both in mass. Two types of overshoot mixing beyond the convective core are considered: according to the $k$-$\omega$ model and according to Herwig (2000). We adopt the overshooting parameter $f_{\rm OV}=0.004$ in Herwig (2000) approach. Finally, we consider three types of their combinations: core overshooting of Herwig (2000) plus complete mixing in the intermediate convection zone (referred to as “OV+Sch”), core overshooting of the $k$-$\omega$ model plus complete mixing in the intermediate convection zone (referred to as “KO+Sch”), and overshooting of the $k$-$\omega$ model in both convective regions (referred to as “KO+KO”).

\begin{figure}
\plotone{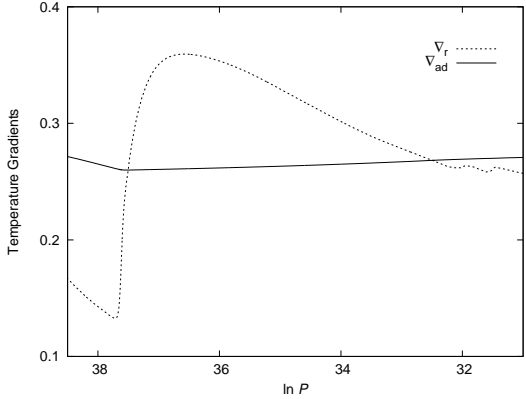}
\figurenum{11}
\caption{Distributions of both radiative and adiabatic temperature gradients for the intermediate convection zone of a KO+KO stellar model in the post main sequence stage. }
\end{figure}

\begin{figure}
\plotone{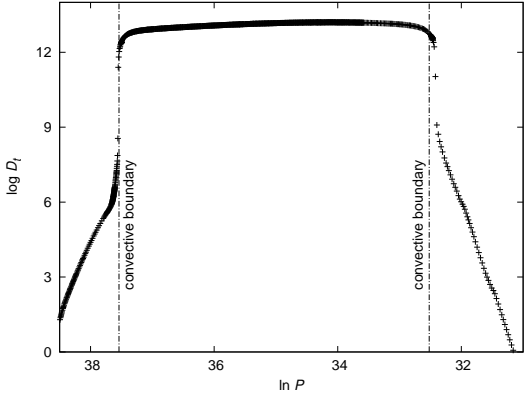}
\figurenum{12}
\caption{Distribution of the turbulent diffusivity for the intermediate convection zone of the KO+KO model in Fig. 11. The vertical lines indicate the Schwarzschild boundaries of the intermediate convection zone.}
\end{figure}

\begin{figure}
\plotone{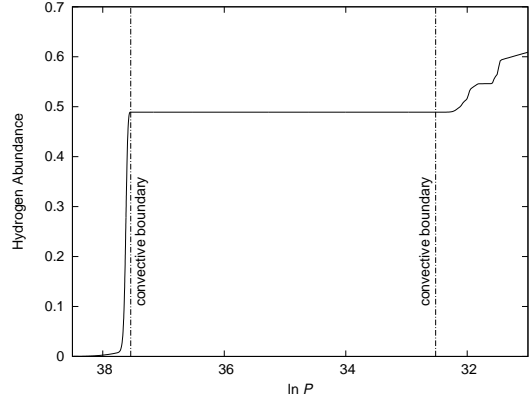}
\figurenum{13}
\caption{Hydrogen profile of the intermediate convection zone for the KO+KO model in Fig. 11. The vertical lines indicate the Schwarzschild boundaries of the intermediate convection zone.}
\end{figure}

The development of the convective core is shown in Figure 9 for the three kinds of models. It can be seen that the overshoot mixing of Herwig (2000) results in the largest convective core during most of time. It should be noted that the development of the convective core is insensitive to the overshooting parameter $f_{\rm OV}$ (our choice is $f_{\rm OV}=0.004$), and different values in the range of $ 0.001 < f_{\rm OV}< 0.004$ give almost the same results. On the other hand, however, the overshoot mixing of the $k$-$\omega$ model results in different convective cores depending on which mixing scheme is used in the intermediate convection zone: the complete mixing leads to a similar convective core as the OV+Sch model, while the overshoot mixing of the $k$-$\omega$ model leads to the smallest convective core (about 0.3$M_{\odot}$ lower in average). It is interesting to notice in this case that different treatments of mixing beyond the hydrogen-burning shell can significantly affect the development of convection within the hydrogen-burning shell. Recently Choi et al. (2016) calculated the MESA stellar models with $f_{\rm OV}=0.016$ both in the main sequence and in the core helium burning stage. The convective cores in their core helium burning models with the same mass will be much bigger than ours. Figure 10 shows the helium profiles for models of the three kinds of mixing schemes at the end of the helium burning phase. It can be seen that the OV+Sch model gives the largest helium depleted core, while the KO+KO model gives the smallest one. 

\begin{figure*}
\plotone{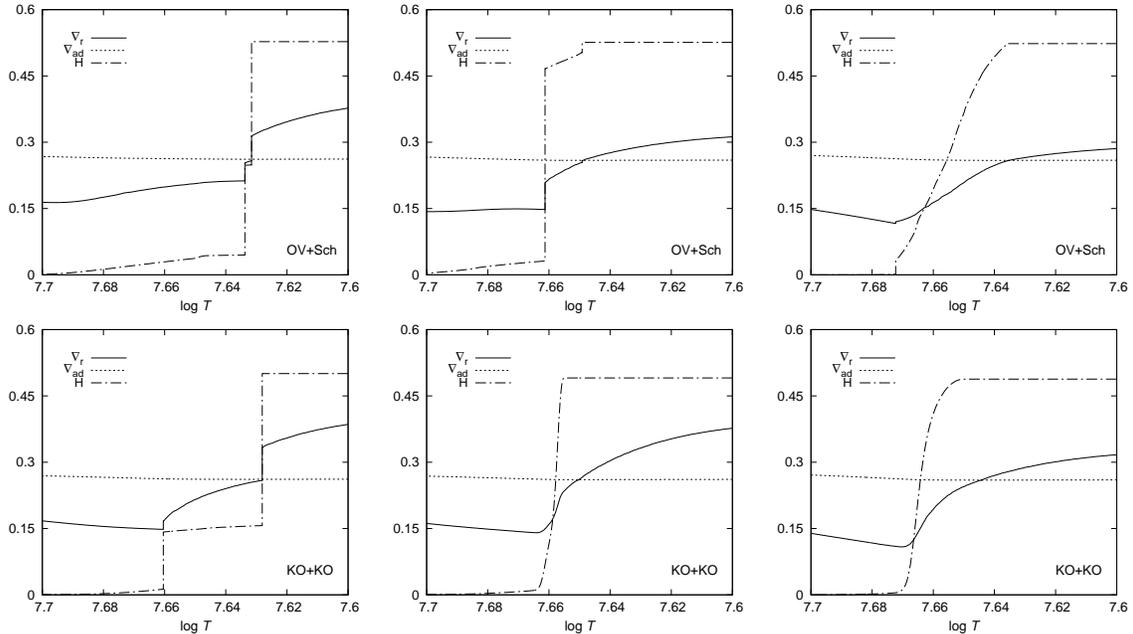}
\figurenum{14}
\caption{Distributions of temperature gradients and hydrogen profiles for OV+Sch (upper panels) and KO+KO (lower panels) stellar models before (left panels), just (middle panels), and after (right panels) the moment when the lower edge of the intermediate convection zone penetrates into the innermost place. The right panels clearly show that the lower edge of the intermediate convection zone is in contact with the hydrogen-burning shell.}
\end{figure*}

\subsection{ The intermediate convection zone }

The profiles of radiative and adiabatic temperature gradients are shown in Figure 11 for the intermediate convection zone beyond the hydrogen-burning shell. It is worth to notice that this intermediate convection zone is fairly wide, occupying a width of about $5H_P$. The corresponding profile of turbulent diffusivity is shown in Figure 12 for the overshoot mixing of the $k$-$\omega$ model. It can be noticed that there are two overshooting regions at both side of the intermediate convection zone: an inner one below and an outer one above. For the inner overshooting region, the turbulent diffusivity drops suddenly to about $10^6{\rm cm}^2{\rm s}^{-1}$ and then decreases exponentially, resulting in an overshooting distance of about $0.8H_P$. For the outer overshooting region, however, the turbulent diffusivity drops only to about $10^9{\rm cm}^2{\rm s}^{-1}$ and then decreases exponentially, leaving an overshooting distance of about $1.2H_P$. The corresponding profile of hydrogen abundance is shown in Figure 13. It can be clearly seen that the hydrogen profile in the two overshooting regions are quite different, showing a sharp step in the inner one but a slow variation in the outer one.

However, the most significant result is that the lower boundary of the intermediate convection zone can touch the upper part of the hydrogen-burning shell. This phenomenon can be clearly seen in Figure 14, which shows the profiles of hydrogen abundance and temperature gradients for the OV+Sch and the KO+KO models around the moment when the lower edge of the intermediate convection zone develops inwardly into the deepest place. It can be seen in the left panels that the lower edge of the intermediate convection zone has not developed inwardly to the innermost place, and also not connected yet with the hydrogen-burning shell at this moment. In the middle panels, the lower edge of the intermediate convection zone has just arrived at the innermost place. For the OV+Sch model, the hydrogen-burning shell has still not been in contact with the lower edge of the intermediate convection zone. For the KO+KO model, however, the upper part of the hydrogen-burning shell has been in contact with the lower edge of the intermediate convection zone due to inclusion of the overshoot mixing. In the right panels, finally, the hydrogen-burning shell has been in contact with the lower edge of the intermediate convection zone for both models, and the overshoot mixing covers most of the hydrogen-burning shell for the KO+KO model. Figure 15 shows the evolution of oxygen profiles in the intermediate convection zone of the KO+KO model. It can be seen clearly that the oxygen content is continuously decreasing, because oxygen is burnt into nitrogen through the ON cycle near the bottom of the intermediate convection zone.

\begin{figure}
\plotone{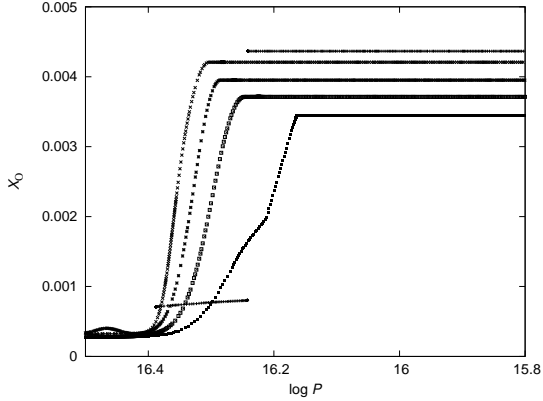}
\figurenum{15}
\caption{Evolution of the oxygen profile in the intermediate convection zone for the KO+KO model. Curves from top to bottom represent oxygen profiles at early to late evolutionary moment.}
\end{figure}

\begin{figure}
\plotone{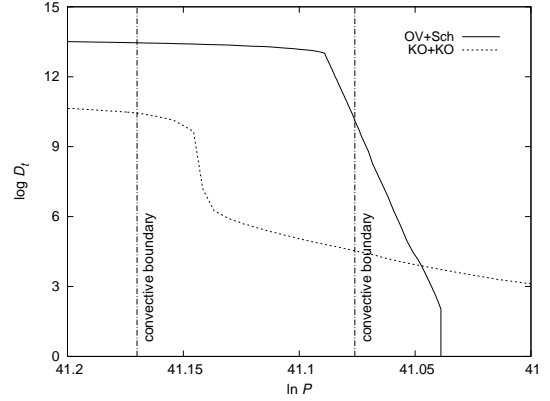}
\figurenum{16}
\caption{Distributions of the turbulent diffusivity for the convective core of OV+Sch and KO+KO model. The left vertical line indicates the Schwarzschild boundary of the convective core for the KO+KO model, while the right one for the OV+Sch model.}
\end{figure}

\begin{figure}
\plotone{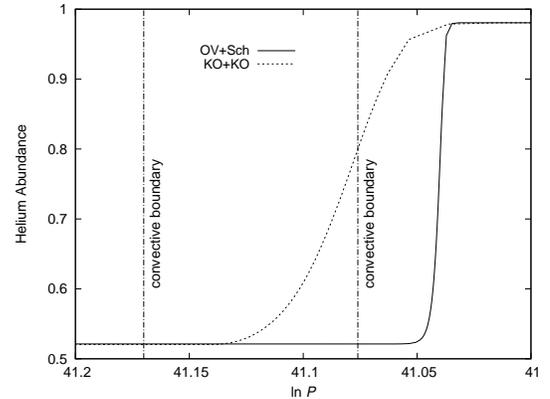}
\figurenum{17}
\caption{Helium profiles near the convective core of OV+Sch and KO+KO model in Fig. 16. The left vertical line indicates the Schwarzschild boundary of the convective core for the KO+KO model, while the right one for the OV+Sch model.}
\end{figure}

\begin{figure}
\plotone{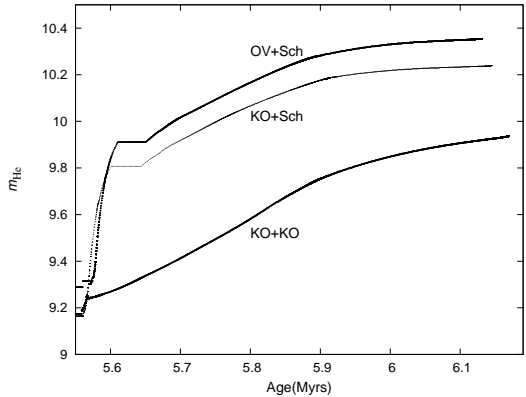}
\figurenum{18}
\caption{Development of the hydrogen-depleted core for the three considered stellar models. The abscissa is the age in unit of Myrs, while the ordinate is the mass of the hydrogen-depleted core in solar mass unit.}
\end{figure}

\subsection{ The convective core }

The profiles of turbulent diffusivity are shown in Figure 16 for the OV+Sch and the KO+KO models during the central helium burning phase. As the size of the convective core is different now for the two considered models, we indicate the boundary of the convective core for both of two models. It can be seen that the values of turbulent diffusivity are big enough for both models to ensure complete mixing within the convective core. In the overshooting regions, however, the turbulent diffusivity decays much rapidly in the OV+Sch model than in the KO+KO model. The corresponding helium profiles are shown in Figure 17. It can be noticed that the structure of the overshooting region is quite different for the two considered models. For the OV+Sch model, the overshooting region is totally about $0.04H_P$, and about half of it is completely mixed. For the KO+KO model, however, the total overshooting region can be about $0.15H_P$, but the complete mixing part is only about $0.03H_P$. This is a convincible indication that the overshooting distance in the core helium burning stage may be significantly smaller than that in the main sequence phase for massive stars.

As mentioned before, the development of the convective core is sensitively dependent on the development of the intermediate convection zone beyond the hydrogen-burning shell. This is because the size of the convective core is closely related to the helium burning efficiency near the stellar center, i.e., the central temperature. On the other hand, the central temperature of the star depends closely on the mass of the hydrogen depleted core, i.e. the location of the hydrogen-burning shell. It can be noticed already in Figure 14 that the lower edge of the intermediate convection zone can penetrate considerably deeper into the stellar interior for the KO+KO model than for the OV+Sch model. Consequently, the hydrogen-burning shell of the KO+KO model will be located at the innermost among the three considered models. Figure 18 shows the development of the hydrogen depleted core for the three considered models. It can be noticed that the KO+KO model has the smallest hydrogen depleted core, and will consequently have the lowest central temperature and the smallest convective core.

\begin{figure}
\plotone{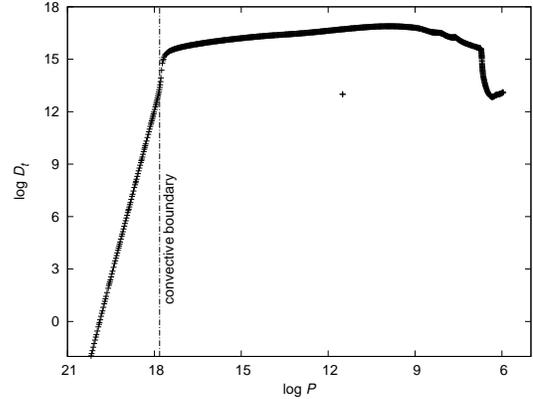}
\figurenum{19}
\caption{The distribution of the turbulent diffusivity for the convective envelop of the KO+KO model. The vertical line shows the Schwarzschild boundary of the base of the convective envelope.}
\end{figure}

\begin{figure}
\plotone{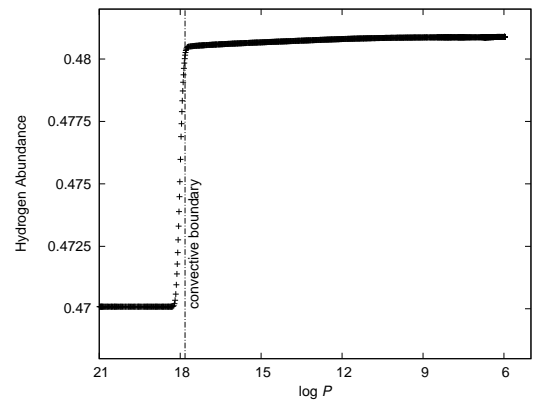}
\figurenum{20}
\caption{The hydrogen profile of the convective envelope for the KO+KO model. The vertical line shows the same Schwarzschild boundary of the base of the convective envelope as in Figure 19. }
\end{figure}

\subsection{ The envelope convection }

During the late stage of the core helium burning, all of the three stars evolve to the red supergiant region in the HR diagram. Then convection develops in the stellar envelope, resulting in the formation of a convective envelope. To compare effects of envelope convection on the evolution of the three stars, we use the standard mixing-length theory in the OV+Sch and KO+Sch models that do not consider the overshooting below the base of the convective envelope, and use the $k$-$\omega$ model in the KO+KO model to include the overshooting below the base of the convective envelope. The typical distribution of the turbulent diffusion coefficient in the KO+KO model is given in Figure 19. It can be seen that turbulent diffusion coefficient is high enough in the convective envelope to ensure a complete mixing up to the stellar surface. Below the base of the convection zone, however, there is an overshooting region with the turbulent diffusion coefficient decaying exponentially inwardly, which is in agreement with the results of many numerical simulations. The overshooting distance is about $2H_P$ as seen in Figure 19, similar to the case of the solar models (Li 2017). The corresponding hydrogen abundance profile is shown in Figure 20. It can be noticed surprisingly that the partially mixing region is very narrow, significantly shorter than the overshooting region of the turbulent diffusion coefficient. This is mainly because that the difference of the hydrogen abundance in and below the convective envelope is also very small (only about 0.01), which is a result of the mass loss due to the stellar wind that makes the star to lose almost all its original hydrogen-abundant envelope and to have an almost chemically homogeneous envelope in the red supergiant stage.

\begin{figure}
\plotone{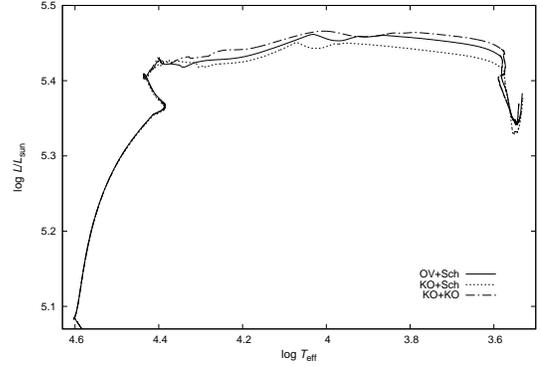}
\figurenum{21}
\caption{The Hertzsprung-Russell diagram for the three considered stellar models. Particularly, the dash-dotted line represents the KO+Semi model in the main sequence stage and the KO+KO model in the subsequent evolution stage.}
\end{figure}

\begin{figure}
\plotone{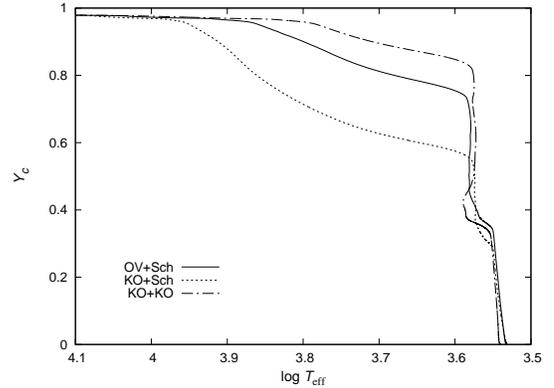}
\figurenum{22}
\caption{Variations of central helium abundance versus the effective temperature for the three stellar models during the central helium burning stage. }
\end{figure}

\subsection{ The evolution of $30M_{\odot}$ stellar models }

The evolution tracks of the three considered models are shown in Figure 21. It should be noted that the dash-dotted line represents the KO+Semi model in the main sequence stage and the KO+KO model in the subsequent evolution stage. It can be seen that the three evolution tracks almost identically coincide with each other in the main sequence stage. This is mainly because we have chosen a specified value of the overshooting parameter $f_{\rm OV}=0.012$ for the OV+Sch model to result in almost the same size of the convective core as in the KO+Sch and the KO+Semi models. It should also be noticed that the detailed hydrogen profile in the chemical gradient region has little effect on the evolution tracks. As soon as hydrogen is depleted in the convective core, the evolution tracks begin to diverge. It can be seen that the OV+Sch model mostly has higher luminosity than the KO+Sch model, because the hydrogen-depleted core of the OV+Sch model is always larger than that of the KO+Sch model. However, the KO+KO model has even higher luminosity than the OV+Sch mode, although its hydrogen-depleted core is the smallest among the three considered models. For the KO+KO model, it plays an important role that the lower edge of the intermediate convection zone can extend into the hydrogen-burning shell. There are two effects that contribute to the stellar luminosity. It can be seen in Figure 14 that the overshooting below the base of the intermediate convection zone helps hydrogen burn at higher temperatures. Besides, the overshoot mixing transports more oxygen simultaneously to the hydrogen burning area, increasing the efficiency of the CNO cycle through the process of oxygen being converted to nitrogen.

Figure 22 shows the variation of the central helium abundance with the effective temperature. It can be noticed that helium ignites at the stellar center when the star has an effective temperature lower than about 10000K, and about 20\% to 40\% of helium is consumed when log $T_{\rm eff} > 3.6$. Among the three considered models, the KO+KO model is the latest to ignite helium at the stellar center and the least to consume helium in higher effective temperature region, while the KO+Sch model is the earliest to ignite helium and the most to consume helium in the higher effective temperature region. It should be noticed that the three models have almost the same age of about 5.59Myr at about log $T_{\rm eff}=4.0$, whereas at about log $T_{\rm eff}=3.6$ the OV+Sch model is about 5.689Myr, the KO+Sch model is about 5.766Myr, and the KO+KO model is about 5.638Myr. 

\begin{figure}
\plotone{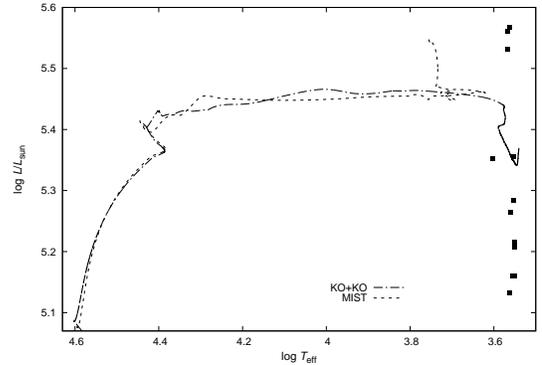}
\figurenum{23}
\caption{Comparisons of the KO+KO model and the MIST model (Choi et al. 2016) with similar composition and the obervations of red supergiants of Milky Way (Levesque et al. 2005). The black squares represent the red supergiants given by Levesque et al. 2005.}
\end{figure}

\begin{figure}
\plotone{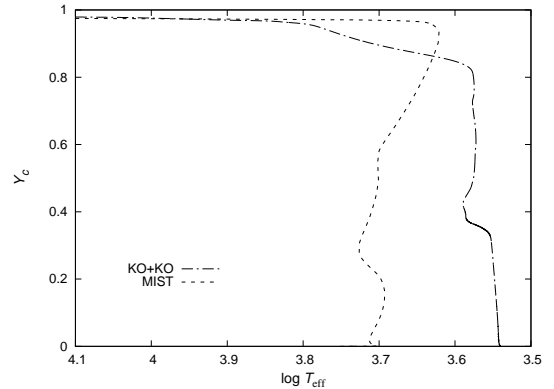}
\figurenum{24}
\caption{Variations of central helium abundance versus the effective temperature for the KO+KO model and the MIST model (Choi et al. 2016). }
\end{figure}

We compare on the HR diagram the KO+KO model and the MIST model (Choi et al. 2016) with similar composition and the observations of red supergiants in the Milk Way (Levesque et al. 2005) in Figure 23. It can be noticed that our result is in good agreement with that of the MIST model (Choi et al. 2016) before the red supergiant stage ($T_{\rm eff}<4100$K). During the red supergiant stage, the KO+KO model can have even lower effective temperatures than the MIST model (Choi et al. 2016), and is in better agreement with the distribution of red supergiants given by Levesque et al. (2005). We also compare the central helium evolution of the KO+KO model with that of the MIST model (Choi et al. 2016) in Figure 24. It can be seen that the KO+KO model ignites the central helium burning at about log $T_{\rm eff}=3.9$ and consumes about 15\% of fuel before it arrives in the red supergiant region, while the MIST model ignites the central helium burning when it arrives almost in the red supergiant region but then evolves back as a yellow supergiant to consume almost its whole fuel.

\section{ Conclusions and Discussions }

Mixing due to convective overshooting and semiconvection are among the most uncertainties in evolution of massive stars. Different approaches give different mixing efficiencies, resulting in great divergence of stellar models. For semiconvection zones, partial mixing due to overstable oscillation proposed by Kato (1966) and double-diffusive approach proposed by Spruit (1992, 2013) are commonly applied in calculations of massive stellar models. For convective overshooting regions, complete mixing over a certain distance beyond the convective boundary (Stothers \& Chin 1985) and an exponentially decaying diffusion outside the convection zone (Herwig 2000) are two commonly adopted approaches. Recently, Li (2012, 2017) developed the $k$-$\omega$ model, which can be used in both convection zones and overshooting regions. We incorporated the $k$-$\omega$ model into MESA, and applied it in calculations of a $30M_{\odot}$ stellar model. We compared the stellar models based on the $k$-$\omega$ model with stellar models based on other approaches to treat the convective overshooting. 

We have found that in the main sequence stage, models based on the $k$-$\omega$ model is almost identical to models with complete mixing in the overshooting region beyond the convective core, and the overshooting in the $k$-$\omega$ model is equivalent to an overshooting distance of about $0.15H_P$. During the late stage of the main sequence, a semiconvection zone is developed near the upper edge of the chemical gradient region. Complete and partial mixing in it give different hydrogen profiles, but those differences do not result in significant difference on the evolution tracks of considered stellar models.

When hydrogen is depleted in the stellar core, hydrogen burning process moves to the surface of the helium core. Meanwhile, an intermediate convection zone is developed beyond the hydrogen-burning shell. We have found that the overshoot mixing below the bottom of the intermediate convection zone can result in two important effects on the stellar evolution. Firstly, such a downward overshooting significantly restricts the size of the hydrogen-depleted core, which will result in a smaller convective core during the central helium burning stage. Secondly, the overshooting penetrates effectively into the hydrogen-burning shell, bringing extra oxygen originally in the intermediate convection zone down to the hydrogen burning area to enhance hydrogen burning efficiency by transferring oxygen into nitrogen through ON cycle.

During the central helium burning stage, a convective core is developed and the mass enclosed in it is increasing with the consumption of helium in the convective core. We have found that the overshooting model based on the $k$-$\omega$ model results in a similar complete mixing region to but a much wider partial mixing region than the overshooting model based on Herwig (2000). About 40\% of helium is burnt near the stellar center in higher effective temperature region (log $T_{\rm eff} > 3.8$), and the inclusion of the overshooting below the bottom of the intermediate convection zone helps to burn more helium in the higher effective temperature region.

In summary, the overshooting distance in the core helium burning stage may be significantly smaller than that in the main sequence phase for massive stars. In addition, the inclusion of overshooting mixing below the bottom of the intermediate convection zone beyond the hydrogen-burning shell is crucial for the evolution of the core helium burning stage.

\acknowledgments 
This work was funded by the NSFC of China (Nos. 11333006, 11703081, 11733008, and 11521303). Hailiang Chen thanks for the fund from Yunnan Province (No. 2017HC018) and Light of West China Program from Chinese Academy of Sciences. Yan Li thank for fruit discussions with Qiansheng Zhang, Tao Wu, Xiaofeng Wang.

\software{MESA(v6596; Paxton et al. 2011, 2013, 2015, 2018)}

{}

\clearpage 

\clearpage

\clearpage


\begin{thebibliography}{}
\bibitem[Briquet et al. (2007)]{br07} Briquet, M., Morel, T., Thoul, A., et al., 2007, \mnras, 381, 1482
\bibitem[Brott et al. (2011)]{br08} Brott, I., de Mink, S.E., Cantiello, M., et al., 2011, \aap, 530, A115
\bibitem[Canuto (1992)]{ca92} Canuto, V.M., 1992, \apj, 392, 218
\bibitem[Canuto (1993)]{ca93} Canuto, V.M., 1993, \apj, 416, 331
\bibitem[Chiosi \& Summa (1970)]{cs70} Chiosi, C. \& Summa, C., 1970, \apss, 8, 478
\bibitem[Choi et al. (2016)]{ch16} Choi, J., Dotter, A., Conroy, C., et al. 2016, \apj, 823, 102
\bibitem[Ding \& Li (2014a)]{dl14a} Ding, C.Y. \& Li, Y., 2014, RAA, 14, 979
\bibitem[Ding \& Li (2014b)]{dl14b} Ding, C.Y. \& Li, Y., 2014, \mnras, 438, 1137
\bibitem[Ekstr\"om et al. (2008)]{ek08} Ekstr\"om, S., Meynet, G., Maeder, A., et al. 2008, \aap, 478, 467
\bibitem[Freytag et al. (1996)]{fr96} Freytag, B., Ludwig, H.-G., Steffen, M., 1996, \aap, 313, 497
\bibitem[Grevesse \& Noels (1993)]{gn93} Grevesse, N., Noels, A., 1993, in Origin and Evolution of the Elements, ed. N. Prantzos, E. Vangioni-Flam, \& M. Casse (Cambridge: Cambridge Univ. Press), 15
\bibitem[Herwig (2000)]{he00} Herwig, F., 2000, \aap, 360, 952
\bibitem[Hossain \& Rodi(1982)]{hr82} Hossain, M.S. \& Rodi, W. 1982, Turbulent Buoyant Jets and Plumes, ed. W. Rodi, The Science and Applications of HMT, 6, 121
\bibitem[Hunter et al. (2008)]{hu08} Hunter, I., Lennon, D.J., Dufton, P.L., et al., 2008, \aap, 479, 541
\bibitem[Kato (1966)]{ka66} Kato, S., 1966, \pasj, 18, 374
\bibitem[Langer (1991)]{la91} Langer, N., 1991, \aap, 243, 155 
\bibitem[Langer (2012)]{la12} Langer, N., 2012, \araa, 50, 107 
\bibitem[Langer \& Maeder (1995)]{lmf95} Langer, N. \& Maeder, A., 1995, \aap, 295, 685
\bibitem[Langer, El Eid \& Fricke (1985)]{lef85} Langer, N., El Eid, M.F. \& Fricke, K.J., 1985, \aap, 145, 179
\bibitem[Langer et al. (1983)]{lsf83} Langer, N., Sugimoto, D., Fricke, K.J., 1983, \aap, 126, 207
\bibitem[Levesque et al. (2005)]{le05} Levesque, E.M., Massey, P., Olsen, K.A.G., et al., 2005, \apj, 628, 973
\bibitem[Li (2012)]{ly12} Li, Y., 2012, \apj, 756, 37
\bibitem[Li (2017)]{ly17} Li, Y., 2017, \apj, 841, 10
\bibitem[Li et al. (2018)]{ly18} Li, Y., Chen, X.H., Xiong, H.R., et al. 2018, \apj, 863, 12
\bibitem[Li \& Yang (2001)]{ly01} Li, Y. \& Yang, J.Y., 2001, ChJAA, 1, 66
\bibitem[Maeder (1987)]{ma87} Maeder, A., 1987, \aap, 178, 159 
\bibitem[Maeder (2009)]{ma09} Maeder, A., 2009, Physics, Formation and Evolution of Rotating Stars (Berlin:Springer) 
\bibitem[Maeder \& Meynet (1987)]{mm87} Maeder, A. \& Meynet, G., 1987, \aap, 182, 243
\bibitem[Maeder \& Meynet (2000)]{mm00} Maeder, A. \& Meynet, G., 2000, \araa, 38, 143 
\bibitem[Maeder \& Meynet (2001)]{mm01} Maeder, A. \& Meynet, G., 2001, \aap, 373, 555
\bibitem[Mermilliod \& Maeder (1986)]{mm86} Mermilliod, J-C. \& Maeder, A., 1986, \aap, 158, 45
\bibitem[Moravveji et al. (2015)]{mo15} Moravveji, E., Aerts, C., P\'apics, P.I., et al. 2015, \aap, 580, A27
\bibitem[Paxton et al. (2011)]{pa11} Paxton, B., Bildsten, L., Dotter, A., et al. 2011, \apjs, 192, 3
\bibitem[Paxton et al. (2013)]{pa13} Paxton, B., Cantiello, M., Arras, P., et al.  2011, \apjs, 208, 4
\bibitem[Paxton et al. (2015)]{pa15} Paxton, B., Marchant, P., Schwarb, J., et al.  2011, \apjs, 220, 15
\bibitem[Paxton et al. (2018)]{pa18} Paxton, B., Schwarb, J., Bauer, E.B., et al. 2018, \apjs, 234, 34
\bibitem[Pope(2000)]{pop} Pope, S.B., 2000, Turbulent Flows (Cambridge: Cambridge University Press)
\bibitem[Schwarzschild \& H\"arm (1958)]{sh58} Schwarzschild, M. \& H\"arm, R., 1958, \apj, 128, 348
\bibitem[Spruit (1992)]{sp92} Spruit, H.C., 1992, \aap, 253, 131
\bibitem[Spruit (2013)]{sp13} Spruit, H.C., 1966, \aap, 552, A76
\bibitem[Stothers \& Chin (1985)]{sc85} Stothers, R.B. \& Chin, C.-W., 1985, \apj, 292, 222
\bibitem[Woosley \& Weaver (1988)]{ww88} Woosley, S.E. \& Weaver, T.A., 1988, Phys. Rep., 163, 79
\bibitem[Woosley et al. (2002)]{whw02} Woosley, S.E., Heger, A., Weaver, T.A., 2002, Rev. Modern Phys., 74, 1015
\bibitem[Xiong (1978)]{xio78} Xiong, D.R., 1978, Chin. Astron., 2, 118
\bibitem[Xiong (1985)]{xio85} Xiong, D.R., 1985, \aap, 150, 133
\bibitem[Xiong (1986)]{xio86} Xiong, D.R., 1986, \aap, 167, 239
\bibitem[Xiong et al. (1997)]{xcd97} Xiong, D.R., Cheng, Q.L., Deng, L.C., 1997, \apjs, 108, 529
\end{thebibliography}
\end{document}